\documentclass[11pt]{revtex4}    
\usepackage{amssymb,epsf}
\usepackage{latexsym}
\begin{document}

\title{Thermodynamic instability of charged dilaton black holes in AdS spaces}
\author{A. Sheykhi$^{1,2}$\footnote{sheykhi@mail.uk.ac.ir}, M. H. Dehghani$^{3}$\footnote{mhd@shirazu.ac.ir} and S. H.
Hendi$^{4}$\footnote{hendi@mail.yu.ac.ir}}

\address{$^1$ Department of Physics, Shahid Bahonar University, P.O. Box 76175-132, Kerman, Iran\\
         $^2$ Research Institute for Astronomy and Astrophysics of Maragha (RIAAM), Maragha,
         Iran\\
         $^3$ Physics Department and Biruni Observatory, Shiraz University, Shiraz 71454,
         Iran\\
         $^4$ Physics Department, College of Sciences, Yasouj University, Yasouj 75914, Iran}

\begin{abstract}
We study thermodynamic instability of a class of
$(n+1)$-dimensional  charged dilatonic spherically symmetric black
holes in the background of anti-de Sitter universe. We calculate
the quasilocal mass of the AdS dilaton black hole through the use
of the subtraction method of Brown and York. We find a Smarr-type
formula and perform a stability analysis in the canonical ensemble
and disclose the effect of the dilaton field on the thermal
stability of the solutions. Our study shows that the solutions are
thermally stable for small $\alpha$, while for large $\alpha$ the
system has an unstable phase, where $\alpha $ is a coupling
constant between the dilaton and matter field.
\end{abstract}
\maketitle

\section{Introduction}
Strong motivation for studying thermodynamics of black holes in
anti-de Sitter (AdS) spaces arises from the correspondence between
the gravitating fields in AdS spacetime and conformal field theory
(CFT) living on the boundary of the AdS spacetime \cite{Witt1}. It
was argued that the thermodynamics of black holes in AdS spaces
can be identified with that of a certain dual CFT in the high
temperature limit \cite{Witt2}. With the AdS/CFT correspondence,
one can gain some insights into thermodynamic properties and phase
structures of strong 't Hooft coupling CFTs by studying
thermodynamics of AdS black holes. It is well-known that the
Schwarzschild black hole in AdS space is thermodynamically stable
for large horizon radius, while it becomes unstable for small
horizon radius. That is, there is a phase transition, named
Hawking-Page phase transition, between the high
temperature black hole phase and low temperature thermal AdS space \cite{Haw}%
. It has been explained by Witten \cite{Witt2} that the
Hawking-Page phase transition of Schwarzschild black holes in AdS
spaces can be identified with confinement/deconfinement transition
of the Yang-Mills theory in the AdS/CFT correspondence. It is
important to note that for the (locally) AdS black holes with zero
or negative constant curvature horizon the Hawking-Page phase
transition does not appear and these black holes are always
locally stable \cite{Birm,Bril,Hendi1}.

On the other side, there has been a renewed interest in studying
scalar coupled solutions of general relativity ever since new
black hole solutions have been found in the context of string
theory. The low energy limit of the string theory leads to the
Einstein gravity, coupled non-minimally to a scalar dilaton field
\cite{Wit1}. The dilaton field couples in a nontrivial way to
other fields such as gauge fields and results into interesting
solutions for the background spacetime \cite{CDB1,CDB2,Hor2}.
These solutions \cite{CDB1,CDB2,Hor2}, however, are all
asymptotically flat. It has been shown that with the exception of
a pure cosmological constant, no dilaton de-Sitter or
anti-de-Sitter black hole solution exists with the presence of
only one Liouville-type dilaton potential \cite{MW}. In the
presence of one or two Liouville-type dilaton potential, black
hole spacetimes which are neither asymptotically flat nor
(anti)-de Sitter [(A)dS] have been studied in different setups
(see e.g \cite{CHM,Cai,Clem,Sheykhi0,Sheykhi1,Hendi2}). With the
combination of three Liouville type dilaton potentials, charged
dilaton black hole/string solutions in the background of (A)dS
spacetime in four \cite{Gao1} and higher dimensional spacetime
\cite{Gao2,Sheykhi2} have been explored. Such potential may arise
from the compactification of a higher dimensional supergravity
model \cite{Gid} which originates from the low energy limit of a
background string theory.

In this paper, we would like to study thermodynamic instability of
asymptotically AdS dilaton black holes in all higher dimensions.
In particular, we shall disclose the effect of the dilaton field
on the thermal stability of the solutions. The motivation for
studying higher dimensional solutions of Einstein gravity
originates from string theory which is believed to be the most
promising approach to quantum theory of gravity in higher
dimensions. In fact, the first successful statistical counting of
black hole entropy in string theory was performed for a
five-dimensional black hole \cite{Stro}. This example provides the
best laboratory for the microscopic string theory of black holes.
Besides, recently it has been realized that there is a way to make
the extra dimensions relatively large and still be unobservable.
This is if we live on a three dimensional surface (brane) embedded
in a higher dimensional spacetime (bulk) \cite{RS,DGP}. In such a
scenario, all gravitational objects such as black holes are higher
dimensional. Furthermore, as mathematical objects, black hole
spacetimes are among the most important Lorentzian Ricci-flat
manifolds in any dimension. One striking feature of the Einstein
equations in more than four dimensions is that many uniqueness
properties holding in four dimensions are lost. For instance,
four-dimensional black holes are known to possess a number of
remarkable features, such as uniqueness, spherical topology,
dynamical stability, and the laws of black hole mechanics. One
would like to know which of these are peculiar to four-dimensions,
and which hold more generally. For a recent review on higher
dimensional black holes see \cite {Emp1}. In the light of all
mentioned above, it becomes obvious that further study on the
thermodynamics of higher-dimensional dilaton black holes in AdS
spaces is of great interest.

This paper is outlined as follows. In Sec. \ref{Field}, we present
the $(n+1) $-dimensional black hole solutions of
Einstein-Maxwell-dilaton theory in AdS background. In Sec.
\ref{Therm}, we obtain the conserved and thermodynamic quantities
of the solutions and verify the validity of the first law of black
hole thermodynamics. We perform a stability analysis in the
canonical ensemble and disclose the effect of the dilaton field on
the thermal stability of the solutions in Sec. \ref{Stab}. We
summarize our results in Sec. \ref{Sum}.

\section{dilaton black holes in AdS spaces\label{Field}}
The action of $(n+1)$-dimensional $(n\geq 3)$
Einstein-Maxwell-dilaton gravity can be written
\begin{equation}
S=-\frac{1}{16\pi }\int d^{n+1}x\sqrt{-g}\left( {R}\text{ }-\frac{4}{n-1}%
(\nabla \Phi )^{2}-V(\Phi )-e^{-4\alpha \Phi /(n-1)}F_{\mu \nu }F^{\mu \nu
}\right) ,  \label{Action}
\end{equation}
where ${R}$ is the scalar curvature{\bf , }$V(\Phi )$ is a potential for the
dilaton field $\Phi $. $\alpha $ is an arbitrary constant governing the
strength of the coupling between the dilaton and the Maxwell field, $F_{\mu
\nu }=\partial _{\mu }A_{\nu }-\partial _{\nu }A_{\mu }$ is the
electromagnetic field tensor and $A_{\mu }$ is the electromagnetic
potential. While $\alpha =0$ corresponds to the usual
Einstein-Maxwell-scalar theory, $\alpha =1$ indicates the
dilaton-electromagnetic coupling that appears in the low energy string
action in Einstein's frame. Varying action (\ref{Action}) with respect to
the gravitational field $g_{\mu \nu }$, the dilaton field $\Phi $ and the
gauge field $A_{\mu }$, yields the following field equations
\begin{equation}
{R}_{\mu \nu }=\frac{4}{n-1}\left( \partial _{\mu }\Phi \partial _{\nu }\Phi
+\frac{1}{4}g_{\mu \nu }V(\Phi )\right) +2e^{-4\alpha \Phi /(n-1)}\left(
F_{\mu \eta }F_{\nu }^{\text{ }\eta }-\frac{1}{2(n-1)}g_{\mu \nu }F_{\lambda
\eta }F^{\lambda \eta }\right) ,  \label{FE1}
\end{equation}
\begin{equation}
\nabla ^{2}\Phi =\frac{n-1}{8}\frac{\partial V}{\partial \Phi }-\frac{\alpha
}{2}e^{-{4\alpha \Phi }/({n-1})}F_{\lambda \eta }F^{\lambda \eta },
\label{FE2}
\end{equation}
\begin{equation}
\nabla _{\mu }\left( e^{-{4\alpha \Phi }/({n-1})}F^{\mu \nu }\right) =0.
\label{FE3}
\end{equation}
We assume the $(n+1)$-dimensional spherically symmetric metric has
the following form
\begin{equation}
d{s}^{2}=-N^{2}(\rho )f^{2}(\rho )dt^{2}+\frac{d\rho ^{2}}{f^{2}(\rho )}%
+\rho ^{2}{R^{2}(\rho )}d\Omega _{n-1}^{2},  \label{metric}
\end{equation}
where $d\Omega _{n-1}^{2}$ denotes the metric of an unit $(n-1)$-sphere and $%
N(\rho )$, $f(\rho )$ and $R(\rho )$ are functions of $\rho $
which should be determined. First of all, the Maxwell equations
(\ref{FE3}) can be integrated immediately, where, for the
spherically symmetric spacetime (\ref{metric}), all the components
of ${F}_{\mu \nu }$ are zero except ${F}_{t\rho }$:
\begin{equation}
{F}_{t\rho }=N(\rho )\frac{qe^{4\alpha {\Phi /(n-1)}}}{\left[ \rho
R(\rho)\right] ^{n-1}},  \label{Ftr}
\end{equation}
where $q$, an integration constant, is the charge parameter of the
black hole.  Our aim here is to construct exact,
$(n+1)$-dimensional black hole solutions of Eqs. (\ref
{FE1})-(\ref{FE3}) for an arbitrary dilaton coupling parameter
$\alpha $. The dilaton potential leading to AdS-like solutions of
dilaton gravity has been found recently \cite{Gao2}. For an
arbitrary value of $\alpha $ in AdS spaces the form of the dilaton
potential ${V}({\Phi })$ in $n+1$ dimensions is chosen as
\begin{eqnarray}
{V}({\Phi }) &=&\frac{2\Lambda}{n(n-2+\alpha
^{2})^{2}}\Bigg\{-\alpha
^{2}\left[ (n+1)^{2}-(n+1)\alpha ^{2}-6(n+1)+\alpha ^{2}+9\right] e^{-4(n-2){%
\Phi /}[(n-1)\alpha ]}  \nonumber  \label{V1} \\
&&+(n-2)^{2}(n-\alpha ^{2})e^{4\alpha {\Phi /}(n-1)}+4\alpha
^{2}(n-1)(n-2)e^{-2{\Phi }(n-2-\alpha ^{2})/[(n-1)\alpha ]}\Bigg
\}.
\end{eqnarray}
Here $\Lambda $ is the cosmological constant. For later
convenience we redefine $\Lambda =-n(n-1)/2l^{2}$, where $l$ is
the AdS radius of spacetime. It is clear the cosmological constant
is coupled to the dilaton in a very nontrivial way. This type of
dilaton potential can be obtained when a higher-dimensional theory
is compactified to four dimensions, including various supergravity
models \cite{Gid}. In the absence of the dilaton field action
(\ref{Action}) reduces to the action of Einstein-Maxwell gravity
with cosmological constant. Using metric (\ref {metric}) and the
Maxwell field (\ref{Ftr}), one can show that the system of
equations (\ref{FE1})-(\ref{FE2}) have solutions of the form
\begin{eqnarray}
N^{2}(\rho ) &=&\left[ 1-\left( \frac{b}{\rho }\right) ^{n-2}\right]
^{-\gamma (n-3)},  \label{frho} \\
f^{2}(\rho ) &=&\Bigg\{\left[ 1-\left( \frac{c}{\rho }\right) ^{n-2}\right] %
\left[ 1-\left( \frac{b}{\rho }\right) ^{n-2}\right] ^{1-\gamma \left(
n-2\right) }-\frac{2\Lambda}{n(n-1)} \rho ^{2}\left[ 1-\left( \frac{b}{\rho }%
\right) ^{n-2}\right] ^{\gamma }\Bigg \}  \nonumber \\
&&\times \left[ 1-\left( \frac{b}{\rho }\right) ^{n-2}\right] ^{\gamma
(n-3)},  \label{grho} \\
{\Phi }(\rho ) &=&\frac{n-1}{4}\sqrt{\gamma (2+2\gamma -n\gamma )}\ln \left[
1-\left( \frac{b}{\rho }\right) ^{n-2}\right] ,  \label{Phirho} \\
R^{2}(\rho ) &=&\left[ 1-\left( \frac{b}{\rho }\right)
^{n-2}\right] ^{\gamma }.  \label{Rrho}
\end{eqnarray}
Here $c$ and $b$ are integration constants and the constant
$\gamma $ is
\begin{equation}
\gamma =\frac{2\alpha ^{2}}{(n-2)(n-2+\alpha ^{2})}.  \label{gamma}
\end{equation}
The charge parameter $q$ is related to $b$ and $c$ by
\begin{equation}
q^{2}=\frac{(n-1)(n-2)^{2}}{2(n-2+\alpha ^{2})}c^{n-2}b^{n-2}.  \label{q}
\end{equation}
According to the Gauss theorem, the electric charge of the black
hole is
\begin{equation}
Q=\frac{1}{4\pi }\int_{\rho \rightarrow \infty }{F}_{t\rho }\sqrt{-{g}}%
d^{n-1}x=\frac{\Omega _{n-1}}{4\pi }q,  \label{Q}
\end{equation}
where $\Omega _{n-1}$ is the volume of the unit $(n-1)$-sphere.
For $\alpha \neq 0$ the solutions become imaginary for $0<\rho <b$
and therefore we should exclude this region from the spacetime.
For this purpose we introduce the new radial coordinate $r$ as
\[
r^{2}=\rho ^{2}-b^{2}\Rightarrow d\rho ^{2}=\frac{r^{2}}{r^{2}+b^{2}}dr^{2}.
\]
With this new coordinate, the above metric becomes
\begin{equation}
d{s}^{2}=-N^{2}(r)f^{2}(r)dt^{2}+\frac{r^{2}dr^{2}}{(r^{2}+b^{2})f^{2}(r)%
}+(r^{2}+b^{2}){R^{2}(r)}d\Omega _{n-1}^{2},  \label{Metric2}
\end{equation}
where the coordinates $r$ assumes the values $0\leq r<\infty $,
and $N(r) $, $f(r)$, $\Phi (r)$ and $R(r)$ are given by Eqs.
(\ref{frho})-(\ref{Rrho}) with replacement $\rho= \sqrt{r^2+b^2}$.

The Kretschmann invariant $R_{\mu \nu \lambda \kappa }R^{\mu \nu
\lambda \kappa }$ and the Ricci scalar $R$ diverge at $r=0$ ($\rho
=b$). Thus, $r=0$ is a curvature singularity. It is worthwhile to note that the scalar field $%
\Phi (\rho)$ and the electromagnetic field $F_{t\rho }$ become
zero as $\rho \rightarrow \infty $. It is also notable to mention
that these solutions are valid for all values of $\alpha $. When
($\alpha =0=\gamma $), these solutions describe the
$(n+1)$-dimensional asymptotically AdS
Reissner-Nordstrom black holes. One should note that the singularity for $%
\alpha \neq 0$ is null, while it is timelike for $\alpha =0$.

The quasilocal mass of the dilaton AdS black hole can be calculated through
the use of the subtraction method of Brown and York (BY) \cite{BY}. Such a
procedure causes the resulting physical quantities to depend on the choice
of reference background. In order to use the BY method the metric should
have the form
\begin{equation}
ds^{2}=-W({\mathcal R})dt^{2}+\frac{d{\cal R}^{2}}{V({\cal R})}+{\cal R}%
^{2}d\Omega ^{2}. \label{Mets}
\end{equation}
Thus, we should write the metric (\ref{metric}) in the above form. To do
this, we perform the following transformation:
\[
{\cal R}=\rho \left[ 1-\left( \frac{b}{\rho }\right) ^{n-2}\right]
^{\gamma /2}.
\]
It is a matter of calculations to show that the metric (\ref{metric})
may be written as (\ref{Mets}) with the following $W$ and $V$:
\begin{eqnarray*}
W({\cal R}) &=&N^2(\rho({\cal R}))f^2(\rho ({\cal R})), \\
V({\cal R)} &=&f^2(\rho ({\cal R}))\left( \frac{d{\cal R}}{d\rho
}\right)
^{2}=\left[1+\frac{1}{2}\left(\gamma(n-2)-2\right)\left( \frac{b%
}{\rho }\right) ^{n-2}\right]^2 \left[ 1-\left( \frac{b%
}{\rho }\right) ^{n-2}\right] ^{(\gamma-2)}f^2(\rho ({\cal R})).
\end{eqnarray*}
The background metric is chosen to be the metric (\ref{Mets}) with
\begin{equation}
W_{0}({\cal R})=V_{0}({\cal R})=f^2_{0}(\rho ({\cal R}))=\left\{
  \begin{array}{ll}

$$1+\frac{\rho^2}{l^2}-\,{\frac {2{\alpha}^{2}b \rho}{{l}^{2}
\left( 1 +{\alpha}^{2} \right) }}+{\frac {{\alpha}^{4}{b}^{2}}{l^2
\left( 1+{\alpha}^{2}
 \right) ^{2}}}\hspace{.4cm}\mathrm{for}\ \ n=3 $$ \\
$$1+\frac{\rho^2}{l^2}-{\frac {{
\alpha}^{2}{b}^{2}}{{l}^{2} \left( 2+{\alpha}^{2} \right)
}}\hspace{2.4cm}\mathrm{for}\ \ n=4 $$\\
$$1+\frac{\rho^2}{l^2} \hspace{4.2cm}\mathrm{for}\ \ n\geq 5$$
\end{array}
\right.
\end{equation}
As you see from above equation, the solutions for $n=3$ and $n=4$
have not "exact" asymptotic AdS behavior. Because of this point,
we cannot use the AdS/CFT correspondence to compute the mass.
Indeed, for $n\geq5$ the metric is exactly asymptotically AdS,
while for $n=3,4$ it is \textit{approximately} asymptotically AdS.
This is due to the fact that if one compute the Ricci scalar then
it is not equal to $-n(n+1)/l^2$. It is well-known that the Ricci
scalar for AdS spacetime should have this value (see e.g.
\cite{Weinberg}). Also, the metrics with $f_0^2(\rho)$ given by
Eq. (17) for $n=3$ and $n=4$ do not satisfy the Einstein equation
with the cosmological constant, while an AdS spacetime should
satisfy Einstein equation with cosmolgical constant, and an
asymptotical AdS should satisfy at infinity. On the other side, at
large $\rho$, the metric behaves as $\rho^2$ and therefore we used
the word \textit{"approximately"} asymptotically AdS.

To compute the conserved mass of the spacetime, we choose a
timelike Killing vector field $\xi $ on the boundary surface
${\cal B}$ of the spacetime (\ref {Mets}). Then the quasilocal
conserved mass can be written as
\begin{equation}
{\cal M}=\frac{1}{8\pi }\int_{{\cal B}}d^{2}\varphi \sqrt{\sigma }\left\{
\left( K_{ab}-Kh_{ab}\right) -\left( K_{ab}^{0}-K^{0}h_{ab}^{0}\right)
\right\} n^{a}\xi ^{b},
\end{equation}
where $\sigma $ is the determinant of the metric of the boundary
${\cal B}$, $K_{ab}^{0}$ is the extrinsic curvature of the
background metric  and $n^{a}$ is the timelike unit normal vector
to the boundary ${\cal B}$. In the context of counterterm method,
the limit in which the boundary ${\cal B}$ becomes infinite
(${\cal B}_{\infty }$) is taken, and the counterterm prescription
ensures that the action and conserved charges are finite. Although
the explicit function $f(\rho({\cal R))}$ cannot be obtained, but
at large $\cal R$ this can be done. Thus, one can calculate the
mass through the use of the above modified Brown and York
formalism as
\begin{equation}
{M}=\frac{\Omega _{n-1}}{16\pi }(n-1)\left[ c^{n-2}+\frac{n-2-\alpha ^{2}}{%
n-2+\alpha ^{2}}b^{n-2}\right] .  \label{Mass}
\end{equation}
In the absence of a non-trivial dilaton field ($\alpha =0$), this expression
for the mass reduces to the mass of the $(n+1)$-dimensional asymptotically
AdS black hole.

\section{Thermodynamics of AdS dilaton black hole}\label{Therm}
\begin{figure}[tbp]
\epsfxsize=7cm \centerline{\epsffile{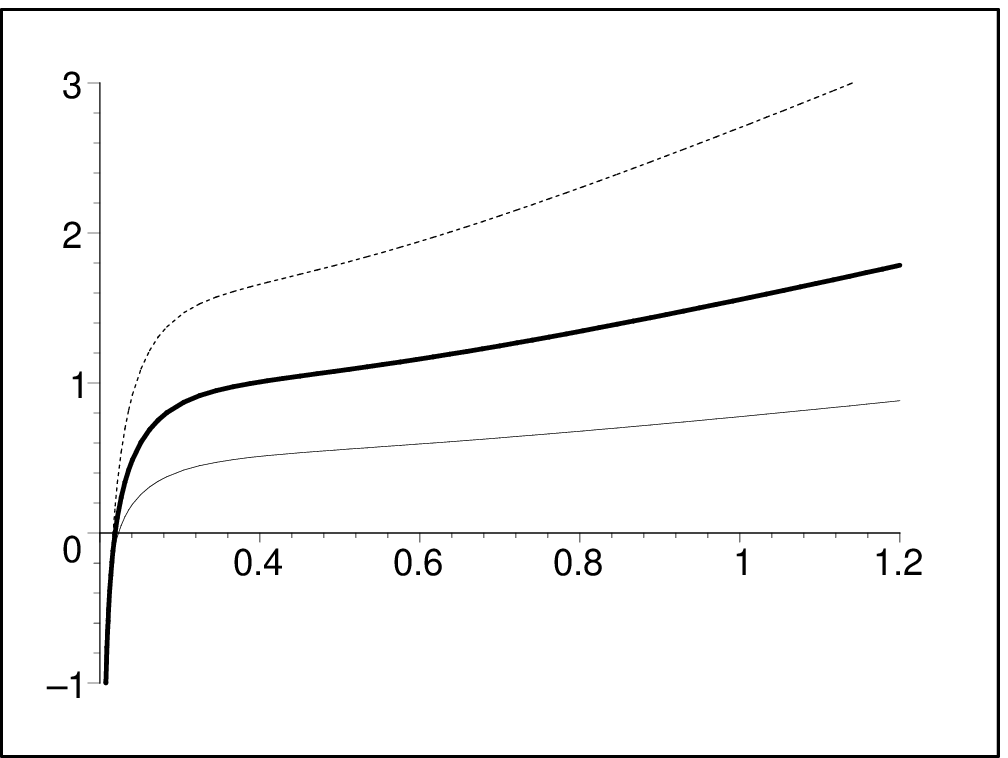}} \caption{T versus
$\rho_{+}$ for $b=0.2$, $l=1$ and $\protect\alpha=0.1,$ $n=4$
(solid line), $n=5$ (bold line), and $n=6$ (dashed line).}
\label{Fig1}
\end{figure}

\begin{figure}[tbp]
\epsfxsize=7cm \centerline{\epsffile{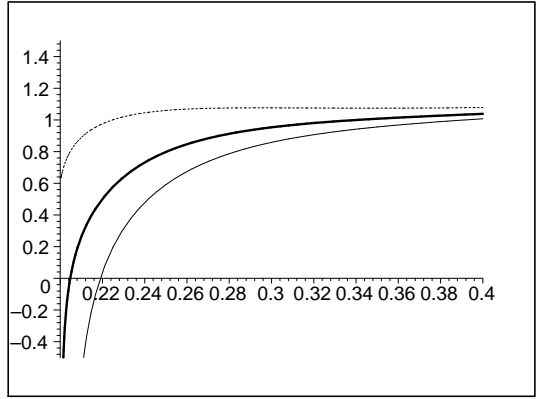}} \caption{T versus
$\rho_{+}$  for  $b=0.2$, $l=1$ and $n=5$. $\protect\alpha=0.1$
(solid line), $\protect\alpha=1$ (bold line), and
$\protect\alpha=2$ (dashed line).} \label{Fig2}
\end{figure}

In this section we intend to study thermodynamics of dilaton black
holes in the background of AdS spaces. The entropy of the dilaton
black hole typically satisfies the so called area law of the
entropy which states that the entropy of the black hole is a
quarter of the event horizon area \cite{Beck}. This near universal
law applies to almost all kinds of black holes, including dilaton
black holes, in Einstein gravity \cite{hunt}. It is a matter of
calculation to show that the entropy of the dilaton black hole is
\begin{equation}
{S}=\frac{\Omega _{n-1}b^{n-1}\Gamma _{+}^{\gamma (n-1)/2}}{4\left( 1-\Gamma
_{+}\right) ^{(n-1)/(n-2)}},  \label{entropy}
\end{equation}
where
\[
\Gamma =1-\left( \frac{b^{2}}{r^{2}+b^{2}}\right) ^{(n-2)/2},
\]
and $\Gamma _{+}=\Gamma (r=r_{+})$ in which $r_{+}$ is the outer
horizon and is related to the parameters $b$, $c$, $\Lambda$ and
$\alpha$. The Hawking temperature of the dilaton black hole on the
outer horizon $r_{+}$, can be calculated using the relation
\begin{equation}
{T}=\sqrt{r^{2}+b^{2}}\left. \frac{\left( N^{2}f^{2}\right) ^{^{\prime }}}{%
4\pi Nr}\right| _{r=r_{+}},
\end{equation}
where a prime stands for the derivative with respect to $r$. One can easily
show that
\begin{eqnarray}\label{Tem}
{T}=\frac{\Lambda b(n-2)\Gamma _{+}^{1-\gamma (n-1)/2}}{2n(n-1)\pi
\left( 1-\Gamma _{+}\right)
^{1/(n-2)}}\Bigg\{\frac{n(n-1)(1-\Gamma _{+})^{2/(n-2)}}{2 \Lambda
b^{2}}-\frac{n\Gamma _{+}^{\gamma (n-1)-1}}{(n-2)}-\frac{\left[
\gamma (n-1)-1\right] (1-\Gamma _{+})}{\Gamma _{+}^{2-\gamma
(n-1)}}\Bigg\}.
\end{eqnarray}
We have shown the behavior of $T$ versus $\rho_{+}$ in various
dimensions in Figs. \ref{Fig1} and \ref{Fig2}. From these figures
we find out that, independent of  the spacetime dimensions, for
small values of $\alpha$ and $\rho_{+}$, the temperature may be
negative ($T<0$).  In this case we encounter a naked singularity.
On the other hand, for an  extremal black hole the temperature is
zero and horizon is degenerate. In this case  $r_{\mathrm{ext}}$
is the positive root of the following equation
\[
\begin{array}{c}
3\Gamma _{\mathrm{ext}}^{2-\gamma (n-1)}\left(1-\Gamma
_{\mathrm{ext}}
\right)^{(4-n)/(n-2)}-\frac{6\Lambda}{n(n-1)} b^{2}\left[ \frac{n\Gamma _{\mathrm{ext}}}{%
(n-2)(1-\Gamma _{\mathrm{ext}})}+\gamma (n-1)-1\right] =0,
\end{array}%
\]
where
\begin{equation}
\Gamma _{\mathrm{ext}}=1-\left( \frac{b}{\sqrt{r_{\mathrm{ext}}^{2}+b^{2}}}%
\right) ^{n-2}%
\end{equation}
Finally, for large values of $\alpha$ it is always positive
provided $\rho_{+}> b$.
Substituting solutions (\ref{frho})-(\ref{Rrho}) in Eq.
(\ref{Ftr}), the electromagnetic field can be simplified as
\begin{equation}\label{Ftrho2}
F_{tr}=\frac{q}{\left( r^{2}+b^{2}\right) ^{\left( n-1\right)
/2}},
\end{equation}
while the corresponding gauge potential $A_{t}$  maybe obtained as
\begin{eqnarray}\label{vectorpot}
A_{t}&=&-\frac{q}{(n-2)\rho^{n-2 }}.
\end{eqnarray}
The electric potential $U$, measured at infinity with respect to
the horizon, is defined by \cite{Cal}
\begin{equation}
U=A_{\mu }\chi ^{\mu }\left| _{r\rightarrow \infty }-A_{\mu }\chi ^{\mu
}\right| _{r=r_{+}},  \label{Pot}
\end{equation}
where $\chi =\partial _{t}$ is the null generator of the horizon. Therefore,
the electric potential may be obtained as
\begin{equation}
U=\frac{q}{(n-2)\rho_{+}^{n-2}},\label{pot}
\end{equation}
where $\rho_{+}^2=r_{+}^2+b^2$. Finally, we check the first law of
thermodynamics for the black hole. In order to do this, we obtain
the mass $M$ as a function of extensive quantities $S$ and $Q$.
Using the expression for the charge, the mass and the entropy
given in Eqs. (\ref{Q}), (\ref{Mass}) and (\ref{entropy}), we can
obtain a Smarr-type formula per unit volume as
\begin{equation}
M(S,Q)=\frac{(n-1)}{16\pi }\left[ \frac{32\pi ^{2}(n-2+\alpha
^{2})Q^{2}b^{2-n}}{(n-1)(n-2)^{2}}+\frac{n-2-\alpha ^{2}}{n-2+\alpha ^{2}}%
b^{n-2}\right] ,  \label{Msmarr}
\end{equation}
where $b=b(Q,S)$. One may then regard the parameters $S$ and $Q$ as a
complete set of extensive parameters for the mass $M(S,Q)$ and define the
intensive parameters conjugate to $S$ and $Q$. These quantities are the
temperature and the electric potential
\begin{eqnarray}
T &=&\left( \frac{\partial {M}}{\partial {S}}\right) _{Q}=\frac{\left( \frac{%
\partial {M}}{\partial {b}}\right) _{Q}\left( \frac{\partial {b}}{\partial {r%
}_{+}}\right) _{Q}}{\left( \frac{\partial {S}}{\partial {r}_{+}}\right)
_{Q}+\left( \frac{\partial {S}}{\partial {b}}\right) _{Q}\left( \frac{%
\partial {b}}{\partial {r}_{+}}\right) _{Q}},  \label{inte1} \\
U &=&\left( \frac{\partial {M}}{\partial {Q}}\right) _{S}+\left(
\frac{\partial {M}}{\partial {b}}\right) _{S}\left( \frac{\partial
{b}}{\partial {Q}} \right) _{S},  \label{inte2}
\end{eqnarray}
where
\begin{eqnarray}
&&\left( \frac{\partial {b}}{\partial {r}_{+}}\right)
_{Q}=-\frac{\left(
\frac{\partial {Z}}{\partial {r}_{+}}\right) _{Q}}{\left( \frac{\partial {Z}%
}{\partial {b}}\right) _{Q}},  \\
&&Z=\left[ 1-\frac{32\pi ^{2}(n-2+\alpha ^{2})Q^{2}}{%
(n-1)(n-2)^{2}r_{+}^{2n-4}}\left( \frac{r_{+}}{b}\right) ^{n-2}\right] \left[
1-\left( \frac{b}{r_{+}}\right) ^{n-2}\right] ^{1-\gamma \left( n-2\right) }
\nonumber \\
&&-\frac{2}{n(n-1)}\Lambda r_{+}^{2}\left[ 1-\left( \frac{b}{r_{+}}\right) ^{n-2}%
\right] ^{\gamma }.
\end{eqnarray}
Straightforward calculations show that the intensive quantities calculated
by Eqs. (\ref{inte1}) and (\ref{inte2}) coincide with Eqs. (\ref{Tem}) and (%
\ref{pot}). Thus, these thermodynamics quantities satisfy the first law of
black hole thermodynamics,
\begin{equation}
dM=TdS+Ud{Q}.
\end{equation}
\section{Stability in the canonical ensemble}\label{Stab}
\begin{figure}[tbp]
\epsfxsize=7cm \centerline{\epsffile{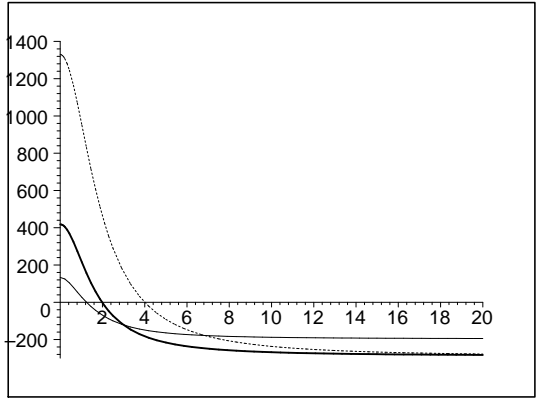}}
\caption{$(\partial ^{2}M/\partial S^{2})_{Q}$ versus $\protect\alpha $ for $%
b=0.2$, $l=1$ and $\rho_{+}=0.4$., $n=5$ (solid line), $n=6$ (bold
line), and $n=7$ (dashed line).} \label{Fig3}
\end{figure}

\begin{figure}[tbp]
\epsfxsize=7cm \centerline{\epsffile{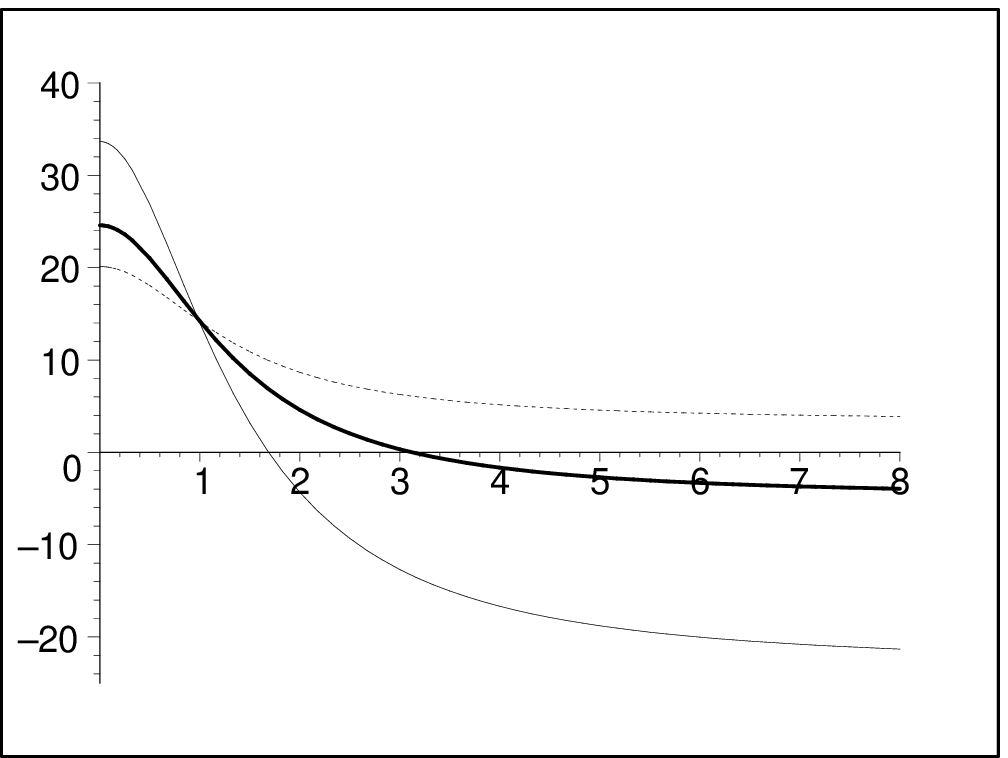}}
\caption{$(\partial ^{2}M/\partial S^{2})_{Q}$ versus $\protect\alpha $ for $%
b=0.2$, $l=1$ and $n=5$. $\rho_{+}=0.5$ (solid line),
$\rho_{+}=0.55$ (bold line), and $\rho_{+}=0.6$ (dashed line).}
\label{Fig4}
\end{figure}

\begin{figure}[tbp]
\epsfxsize=7cm \centerline{\epsffile{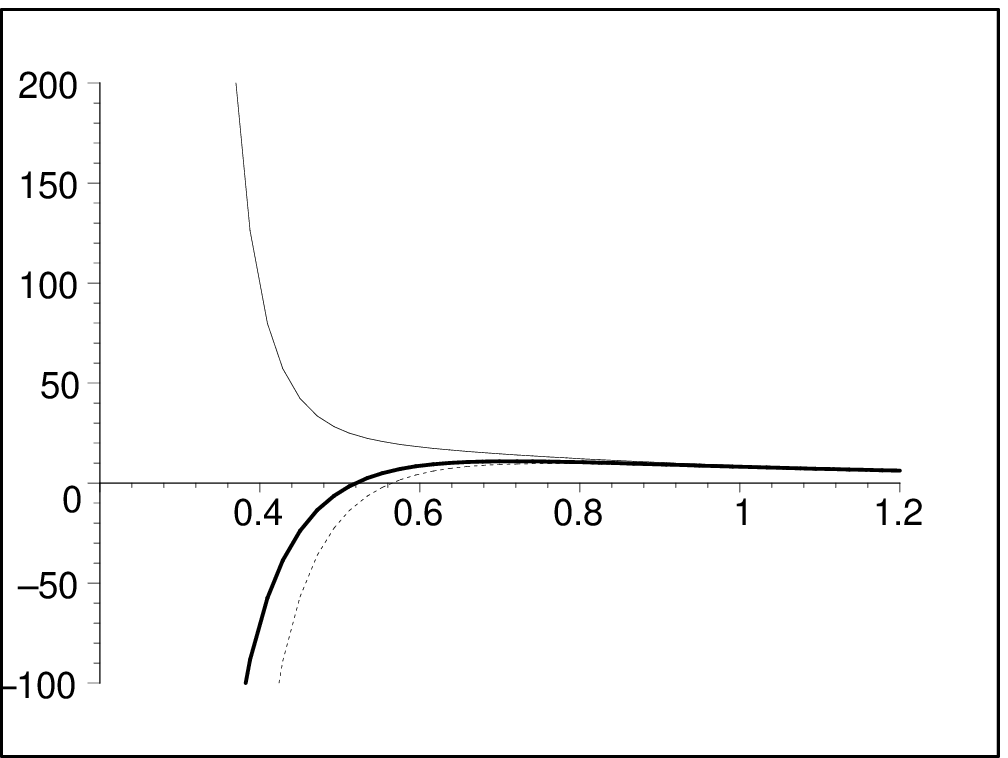}}
\caption{$(\partial ^{2}M/\partial S^{2})_{Q}$ versus $\rho_{+}$
for $b=0.2$, $l=1$ and $n=5$. $\protect\alpha=0.5$ (solid line),
$\protect\alpha=2$ (bold line), and $\protect\alpha=5$ (dashed
line).} \label{Fig5}
\end{figure}

\begin{figure}[tbp]
\epsfxsize=7cm \centerline{\epsffile{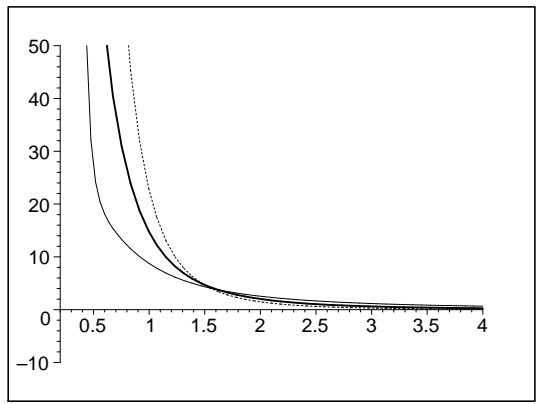}}
\caption{$(\partial ^{2}M/\partial S^{2})_{Q}$ versus $\rho_{+}$
for $b=0.2$, $l=1$ and $\protect\alpha=0.5$. $n=5$ (solid line),
$n=6$ (bold line), and $n=7$ (dashed line).} \label{Fig6}
\end{figure}

Finally, we study the thermal stability of the solutions in the
canonical ensemble. In particular, we will see that the scalar
dilaton field makes the solution unstable. The stability of a
thermodynamic system with respect to small variations of the
thermodynamic coordinates is usually performed by analyzing the
behavior of the entropy $ S(M,Q)$ around the equilibrium. The
local stability in any ensemble requires that $S(M,Q)$ be a
concave function of the intensive variables. The stability can
also be studied by the behavior of the energy $M(S,Q)$ which
should be a convex function of its extensive variable. Thus, the
local stability can in principle be carried out by finding the
determinant of the Hessian matrix of $M(S,Q)$ with respect to its
extensive variables $X_{i}$,
$\mathbf{H}_{X_{i}X_{j}}^{M}=[\partial ^{2}M/\partial
X_{i}\partial X_{j}]$ \cite{Cal,Gub}. In our case the mass $M$ is
a function of entropy and charge. The number of thermodynamic
variables depends on the ensemble that is used. In the canonical
ensemble, the charge is a fixed parameter and therefore the
positivity of the $(\partial ^{2}M/\partial S^{2})_{Q}$ is
sufficient to ensure local stability. In Fig. \ref{Fig3} we show
the behavior of the $(\partial ^{2}M/\partial S^{2})_{Q}$\ as a
function of the coupling constant parameter $\alpha $ for
different value of $n$. This figure shows that there exists an
upper limit for $\alpha $, named $\alpha_{\mathrm{\max }}$, for
which $(\partial ^{2}M/\partial S^{2})_{Q}$ is negative provided
$\alpha
>\alpha _{\mathrm{\max }}$ and positive otherwise. That is the
black hole solutions are unstable for large values of $\alpha $.
It is important to note that $\alpha _{\mathrm{\max }}$ depends on
the parameters $b$, $r_{+}$ and the dimensionality of spacetime
(see Fig. \ref{Fig4}). On the other hand, Figs. \ref{Fig5} and
\ref{Fig6} show the behavior of the $(\partial ^{2}M/\partial
S^{2})_{Q}$ as a function of the $\rho_{+}$ for different value of
coupling constant parameter $\alpha $ and
different value of $n$. These figures show that for a fixed value of $%
\alpha $ the solution is thermally stable for $\rho_{+}>b$ in 5,6
and 7 dimension provided $\alpha<\alpha_{\mathrm{\max }}$. It is
also easy to plot these figures for arbitrary $n$ and generalize
the conclusions for higher dimensions.

In comparison with the asymptotically AdS black holes of Einstein
gravity, which have a small unstable phase, the stability phase
structure of the black holes in Einstein-Maxwell-dilaton gravity
shows that the dilaton field crucially changes the stability phase
structure.
\section{summary and Discussion}\label{Sum}
Thermodynamics of black holes in AdS spaces have been the subject
of much recent interest. This is primarily due to their relevance
for the AdS/CFT correspondence. It was argued that the
thermodynamics of black holes in AdS spaces can be identified with
that of a certain dual CFT in the high temperature limit. In this
paper, we considered asymptotically AdS black holes in
$(n+1)$-dimensional Einstein-Maxwell-dilaton gravity. We computed
the charge, mass, temperature, entropy and electric potential of
the AdS dilaton black holes and found that these quantities
satisfy the first law of black hole thermodynamics. We also
obtained a Smarr-type formula, $M(S,Q)$, and performed a stability
analysis in the canonical ensemble by considering $(\partial
^{2}M/\partial S^{2})_{Q}$ for the charged black hole solutions in
$(n+1)$ dimensions and showed that there is no Hawking-Page phase
transition in spite of charge of the black holes for small $\alpha
$, while the solutions have an unstable phase for large values of
$\alpha $. Indeed, for fixed values of the metric parameters, we
found that there exists a maximum value of $\alpha$ for which the
solutions are thermally unstable if $\alpha
>\alpha _{\mathrm{\max }}$, where $\alpha _{\mathrm{\max }}$
depends on the dimensionality of the spacetime and the metric
parameters $b$ and $r_{+}$. This phase behavior shows that although
there is no Hawking-Page transition for black object whose horizon
is diffeomorphic to $\mathbb{R}^{p}$ for small $\alpha$ and
therefore the system is always in the high temperature phase
\cite{Witt2}, but in the presence of dilaton with $\alpha
>\alpha _{\mathrm{\max }}$ the black hole solutions have some unstable
phases.

Finally, we would like to mention that although charged AdS black
holes in dilaton gravity are thermodynamically unstable for large
values of dilaton coupling constant $\alpha$, it is worthwhile to
examine the dynamical (gravitational) instability of these dilaton
black holes. This is due to the fact that there are black holes in
Einstein gravity which are thermodynamically unstable, while they
are dynamically stable \cite{Kon}. However, there may be some
correlations between the dynamic and thermodynamic instability  of
black hole solutions in dilaton gravity \cite{Gub}.
\acknowledgments{We thank the anonymous referees for constructive
comments. This work has been supported by Research Institute for
Astronomy and Astrophysics of Maragha, Iran.}


\begin{references}
\bibitem{Witt1}  E. Witten, Adv. Theor. Math. Phys. {\bf 2}, 253 (1998);%
\newline
J. M. Maldacena, Adv. Theor. Math. Phys. {\bf 2}, 231 (1998).

\bibitem{Witt2}  E. Witten, Adv. Theor. Math. Phys. {\bf 2}, 505 (1998).

\bibitem{Haw}  S. W. Hawking and D. N. Page, Commun. Math. Phys. {\bf 87},
577 (1983).

\bibitem{Birm}  D. Birmingham, Class. Quant. Grav. {\bf 16}, 1197 (1999).

\bibitem{Bril}  D. R. Brill, J. Louko and P. Peldan, Phys. Rev. D {\bf 56},
3600 (1997);\newline
A. Chamblin, R. Emparan, C. V. Johnson and R. C. Myers, Phys. Rev. D {\bf 60}%
, 064018 (1999).

\bibitem{Hendi1}  M. H. Dehghani and S. H. Hendi, Int. J. Mod. Phys. D {\bf %
16}, 1829 (2007);\\ M. H. Dehghani, N. Alinejadi and S. H. Hendi,
Phys. Rev. D {\bf 77}, 104025 (2008).

\bibitem{Wit1}  M. B. Green, J. H. Schwarz and E. Witten, {\it Superstring
Theory}, \\ (Cambridge University Press, Cambridge 1987).

\bibitem{CDB1}  G. W. Gibbons and K. Maeda, Nucl. Phys. B {\bf 298}, 741
(1988); \\ T. Koikawa and M. Yoshimura, Phys. Lett. B {\bf 189},
29 (1987);\\ D. Brill and J. Horowitz, Phys. Lett. B {\bf 262},
437 (1991).

\bibitem{CDB2}  D. Garfinkle, G. T. Horowitz and A. Strominger, Phys. Rev. D
{\bf 43}, 3140 (1991);\newline
R. Gregory and J. A. Harvey, Phys. Rev. D {\bf 47}, 2411 (1993).

\bibitem{Hor2}  G. T. Horowitz and A. Strominger, Nucl. Phys. B {\bf 360},
197 (1991).

\bibitem{MW}  S. J. Poletti, D. L. Wiltshire, Phys. Rev. D {\bf 50}, 7260
(1994);\newline
S. J. Poletti, J. Twamley and D. L. Wiltshire, Phys. Rev. D {\bf 51}, 5720
(1995); \newline
S. Mignemi and D. L. Wiltshire, Phys. Rev. D {\bf 46,} 1475 (1992).

\bibitem{CHM}  K. C. K. Chan, J. H. Horne and R. B. Mann, Nucl. Phys. B {\bf %
447}, 441 (1995).

\bibitem{Cai}  R. G. Cai, J. Y. Ji and K. S. Soh, Phys. Rev D {\bf 57}, 6547
(1998).

\bibitem{Clem}  G. Clement, D. Gal'tsov and C. Leygnac, Phys. Rev. D {\bf 67}%
, 024012 (2003).

\bibitem{Sheykhi0}  A. Sheykhi, M. H. Dehghani, N. Riazi, Phys. Rev. D {\bf %
75}, 044020 (2007); \newline
A. Sheykhi, M. H. Dehghani, N. Riazi and J. Pakravan Phys. Rev. D {\bf 74},
084016 (2006);\newline
A. Sheykhi, N. Riazi, Phys. Rev. D {\bf 75}, 024021 (2007).

\bibitem{Sheykhi1}  A. Sheykhi, Phys. Rev. D {\bf 76}, 124025 (2007);
\newline
A. Sheykhi, Phys. Lett. B {\bf 662}, 7 (2008).

\bibitem{Hendi2}  M. H. Dehghani, J. Pakravan and S. H. Hendi, Phys. Rev. D
{\bf 74}, 104014 (2006); \\ M. H. Dehghani, et al., J. Cosmol.
Astropart. Phys. {\bf 02}, 020 (2007);\\ S. H. Hendi, J. Math.
Phys. {\bf 49}, 082501 (2008).

\bibitem{Gao1}  C. J. Gao, S. N. Zhang, Phys. Rev. D {\bf 70}, 124019 (2004);%
\newline
A. Sheykhi, Phys. Rev. D {\bf 78}, 064055 (2008);\newline
A. Sheykhi, Phys. Lett. B {\bf 672,} 101 (2009).

\bibitem{Gao2}  C. J. Gao, S. N. Zhang, Phys. Lett. B {\bf 605}, 185 (2005);
\newline
C. J. Gao, S. N. Zhang, Phys. Lett. B {\bf 612} 127 (2005).

\bibitem{Sheykhi2}  A. Sheykhi, M. Allahverdizadeh, Phys. Rev. D {\bf 78},
064073 (2008);\newline
A. Sheykhi, M. Allahvedizadeh, Gen. Rel. Grav. in press (2009); \newline
A. Sheykhi, M. M. Yazdanpanah, Phys. Lett. B {\bf 679,} 311 (2009).

\bibitem{Gid}  S. B. Giddings, Phys. Rev. D {\bf 68}, 026006 (2003); \newline
E. Radu, D. H. Tchrakian, Class. Quantum Grav. {\bf 22}, 879 (2005).

\bibitem{Stro}  A. Strominger and C. Vafa, Phys. Lett. B {\bf 379}, 99
(1996).

\bibitem{RS}  L. Randall, R. Sundrum, Phys. Rev. Lett. {\bf 83}, 3370 (1999);%
\newline
L. Randall, R. Sundrum, Phys. Rev. Lett. {\bf 83}, 4690 (1999).

\bibitem{DGP}  G. Dvali, G. Gabadadze, M. Porrati, Phys. Lett. B {\bf 485},
208 (2000);\newline
G. Dvali, G. Gabadadze, Phys. Rev. D {\bf 63}, 065007 (2001).

\bibitem{Emp1}  R. Emparan and H. S. Reall, arXive:0801.3471.

\bibitem{BY}  J. Brown and J. York, Phys. Rev. D {\bf 47}, 1407 (1993);%
\newline
J.D. Brown, J. Creighton, and R. B. Mann, Phys. Rev. D {\bf 50}, 6394 (1994).

\bibitem{Weinberg} S. Weinberg, \textit{Gravitation and
cosmology,} (John Wiley 1972, Chapter 15).

\bibitem{Beck}  J. D. Beckenstein, Phys. Rev. D {\bf 7}, 2333 (1973);\newline
S. W. Hawking, Nature (London) {\bf 248}, 30 (1974);\newline
G. W. Gibbons and S. W. Hawking, Phys. Rev. D {\bf 15}, 2738 (1977).

\bibitem{hunt}  C. J. Hunter, Phys. Rev. D {\bf 59}, 024009 (1999);\newline
S. W. Hawking, C. J Hunter and D. N. Page, Phys. Rev. D {\bf 59}, 044033
(1999);\newline
R. B. Mann Phys. Rev. D {\bf 60}, 104047 (1999).

\bibitem{Cal}  M. Cvetic and S. S. Gubser, J. High Energy Phys. {\bf 04},
024 (1999); \newline
M. M. Caldarelli, G. Cognola and D. Klemm, Class. Quant. Gravit. {\bf 17},
399 (2000).

\bibitem{Gub}  S. S. Gubser and I. Mitra, J. High Energy Phys. {\bf 08}, 018
(2001).

\bibitem{Kon}  R. A. Konoplya and A. Zhidenko, Phys. Rev. D {\bf 78}, 104017
(2008); \newline
D. Birmingham and S. Mokhtari, Phys. Rev. D {\bf 76}, 124039 (2007).

\end{references}
\end{document}